\documentclass[twocolumn]{article}
\usepackage{setspace}
\usepackage{graphicx}
\usepackage[super]{natbib}
\usepackage{amssymb, latexsym}
\usepackage{amsmath} 
\usepackage{amsthm}
\usepackage{bm}
\usepackage[mathscr]{eucal}
\usepackage{enumerate}
\usepackage{multirow} 
\usepackage[center]{caption}

\usepackage{geometry}

\usepackage{authblk}
\usepackage{abstract}

\title{Cochlear detection of double-slip motion in cello bowing}

\author{Rolf Bader}

\affil{
	Institute of Systematic Musicology\\ 
	University of Hamburg\\ 
	Neue Rabenstr. 13, 20354 Hamburg, Germany, R\_Bader@t-online.de}

\author{Robert Mores}

\affil{University of Applied Science\\
	Faculty of Design, Media, Information\\
	Finkenau 35\\
	22081 Hamburg, Germany,  Robert.Mores@haw-hamburg.de}

\begin{document}
	
\twocolumn[
	\begin{@twocolumnfalse}
		\maketitle

\begin{abstract}
A double-slip motion of a cello sound is investigated experimentally with a bowing machine and analyzed using a Finite-Difference Time Domain (FDTD) cochlear model. A double-slip sound is investigated. Here the sawtooth motion of normal bowing is basically present, but within each period the bow hair tears off the strings once more within the period, resulting in a blurred sound. This additional intermediate slip appears around the middle of each period and drifts temporally around  while the sound progresses. When the double-slip is perfectly in the middle of one period the sound is that of a regular sawtooth motion. If not, two periodicities are present around double the fundamental periodicity, making the sound arbitrary. Analyzing the sound with a Wavelet-transform, the expected double-peak of two periodicities around the second partial cannot be found. Analyzing the tone with a cochlear FDTD model including the transfer of mechanical energy into spikes, the doubling and even more complex behaviour is perfectly represented in the Interspike Interval (ISI) of two adjacent spikes. This cochlear spike representation fits perfectly to an amplitude peak detection algorithm, tracking the precise time point of the double-slip within the fundamental period. Therefore the ear is able to detect the double-slip motion right at the transition from the basilar membrane motion into electrical spikes.  
\end{abstract}

\end{@twocolumnfalse}
]

\vspace{2cm}

\section{Introduction}

Bowing a string is a highly nonlinear process, where the bow is adjacently sticking at and slipping over the string \cite{Schelleng1973} \cite{Raman 1918} \cite{McIntyre and Woodhouse 1979} \cite{Cremer1985} \cite{Gueth 1996} \cite{Woodhouse and Galluzzo 2004}. When playing a cello or violin within a normal range of bowing pressure and velocity, a regular Helmholtz or sawtooth waveform is present. The relation between bowing pressure and the position of the bow on the string necessary to result in such a Helmholtz motion is a region in a so-called Schelleng-diagram \cite{Schelleng1973}. Here the stability conditions were considered \cite{Weinrich and Causse 1991} to reach such a normal sawtooth motion. The mechanism was found to be a nonlinear \cite{Mueller and Lauterborn 1996} and self-organizing \cite{Bader2013} system. 

The Helmholtz motion is only one from a variety of possible waveforms or regimes of this system. Transients or tone onsets are often much more complex \cite{Woodhouse and Schumacher 1995} and found to consist of many initial impulses \cite{Gueth 1980} \cite{Gueth 1995}. When calculating a fractal correlation dimension of time series, a perfect harmonic overtone spectrum results in a fractal dimension of one or slightly higher, taking the quasi steady-state motion into account. Still such transients have fractal dimension numbers of two or higher, representing the transient complexity \cite{Bader2013}, depending on attack and articulation. Therefore a multiple of other regimes next to the regular sawtooth motion are possible and part of performance practice. Some are akin to the sawtooth motion and some are more complex including scratch sounds. Simulations of the system \cite{McIntyre et al. 1981} \cite{Woodhouse 2003} \cite{Bader2005} among others show details of such complex behaviour.

Still next to transients, quasi steady-state waveforms of bowing may also be more complex than a regular Helmholtz motion. Complex waveforms may contain subharmonics \cite{Kimura 1999} \cite{Bader2013}, instabilities \cite{McIntyre et al. 1981} \cite{Duffour and Woodhouse 2004a} \cite{Duffour and Woodhouse 2004b} or the so-called double-slip motion \cite{Woodhouse1995}, producing a sound known as 'surface' sound, as it is perceived as slightly blurred or arbitrary. This double-slip motion is caused an additional intermediate slip (string tears off the bow) in between such two major slips, which define the fundamental period of the sound. The intermediate and less prominent slip in the middle is a secondary slip. Therefore within one fundamental period there are two slips, the waveform shows a double-slip.

Fig. 1 shows a piezo accelerometer recording time series of a cello tone played by the bowing machine discussed below. The top plot shows an excerpt of 50 adjacent periods of this tone with a fundamental frequency of 148 Hz which therefore has a periodicity of 6.75 ms. The waveform starts with a maximum acceleration at the point where the bow tears-off the string. The following waveform results from the vibrations of the bridge and the string. Still around the middle of the period a second acceleration peak appears which is larger than the previous ones, and thus represents the mentioned secondary, intermediate slip within the fundamental period. The top plot of Fig. 1 shows 50 adjacent periods of the sound where a high consistency of this second slip is clearly visible. Still the relative time point within one period where double-slip occurs may change during the played tone duration. In the upper middle plot again 50 adjacent periods of the same tone are shown starting right at the end of the time frame shown in the top plot. Again there is a high consistency of the waveform over these 50 periods, however, the instant of time for the intermediate slip drifts forth and back. This is clearly visible when plotting both time frames, that of the top and that of the upper middle plot in one plot as shown in the lower middle plot which now shows 100 periods of the played tone. While the beginning of the waveform is very much the same in both time frames, the waveform differs considerable in the middle of the period, as the time point of the double-slip varies. For comparison the lowest plot shows 50 adjacent periods of a regular sawtooth motion. These are likewise stable but do not show a double-slip around the middle of the period.

The timbre of the sound is not that of a regular cello tone but sounds blurred with some arbitrariness and is therefore called a 'surface' tone. Such sounds are used very often as extended technique in modern string music, especially in film or contemporary classical music. 

\begin{figure}
	\centering
	\includegraphics[width=0.8\linewidth]{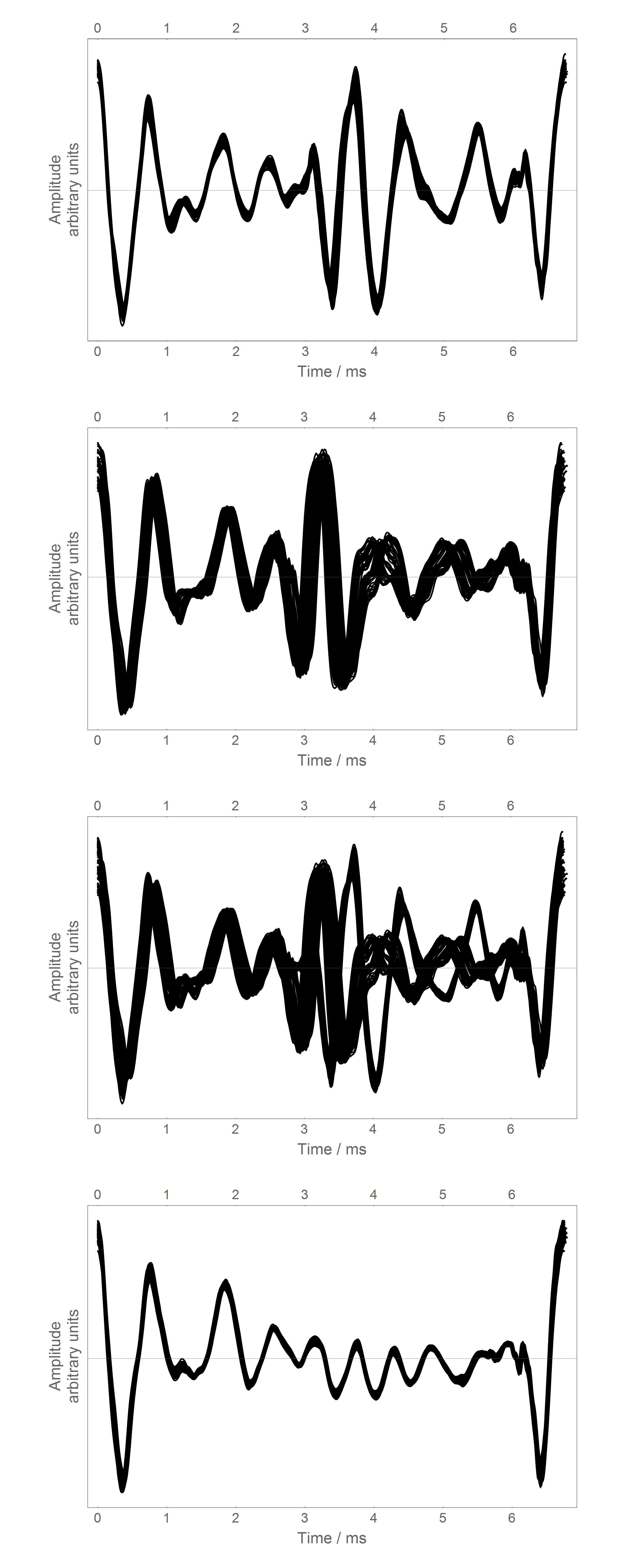}
	\caption{Piezo accelerometer recordings at a cello bridge of a double-slip motion taken from one tone of 148 Hz played by the bowing machine used in this investigation. Top: 50 adjacent periods of a time interval, Upper middle: 50 adjacent periods of a time interval following that of the top plot, Lower middle: top and upper middle plot combined. The double-slip is visible as a peak at about the middle of the period. Top and upper middle plots show high consistency of the time point of the double-slip within the time frame, lower middle plot shows that the time point of the double-slip is different between the time frames. The lowest plot shows the time series for a regular sawtooth motion. No double slip peak in the middle of the period is present there.}
	\label{fig:fig1}
\end{figure}

Now these terms are a semantic representation of the timbre perception of this sound. The present paper is comparing two analyzing techniques for such sounds, that of a Fourier analysis in the frequency domain and that of a cochlear physical model analyzing the sound in the time domain. It will be shown that the cochlear time-domain representation is able to capture the physical presence of a double-slip as a bifurcation with the fundamental frequency being roughly doubled. On the contrary, the frequency-domain represents the double-slip in form of a variation of amplitude for higher frequencies when compared with a regular sawtooth motion. Such interpretation in the sense of timbre alternations is not adequately representing the observed arbitrariness in the time-domain.

The interspike intervals (ISI) of the common denominator of a harmonic sound corresponds to the pitch of this sound. How pitch perception is realized in the brain is under dabate. Pitch perception has been investigated in terms of the cochlear and neural networks, where frequency and time or interspike theories have been proposed. The frequency models correspond to the place theory of frequency representation on the cochlear along cochlear tuning curves, while the time theories are based on the temporal discharge pattern of neurons (see \cite{Cariani2001}, \cite{Lyon1996} for reviews). Licklider \cite{Licklider1956} proposed an autocorrelation model of pitch perception where the time period between peaks of the autocorrelation function of complex harmonic sounds serves as pitch. Schouten \cite{Schouten1962} and later Terhardt proposed a theory of residuals perception in the same direction pointing to the perceptual phenomenon of perceiving a pitch at a frequency where no frequency component is present but where the common denominator of the other frequencies of the harmonic spectrum are placed. Lyon and Shamma \cite{Lyon1996} propose a temporal theory of pitch perception also discussing musical intervals of simple integer rations (such as 3:2 of the musical fifth or 4:3 of the fourth, etc.). So although there is no conclusive proof that ISI of a common denominator corresponds to the perceived pitch, most theories are based on this assumption and therefore it is used in the following.

For extracting a cochleogram from a time series, gammatone filter banks are most often used (for reviews see \cite{Patterson1995}, \cite{Cariani1999}). Here the transformation from mechanical waves on the BM to spikes uses Fourier analysis to model the best frequency, spectral filters to model fluid-cilia coupling and hair cell movement and nonlinear compression to address logarithmic sound pressure level (SPL) modulation. These gammatone filter banks do not use physical modeling of the BM or the lymph and are based on signal processing analogies. Although they cover many aspects of mechanical-to-electrical transfer they are not able to detect synchronization effects as discussed here.

Many physical models of the cochlear have been proposed differing according to the problem addressed in the respective study, among others like phase-consistency \cite{Steele1979} and frequency dispersion \cite{Ramamoorthy2010}, lymph hydrodynamics \cite{Mammano1992}, active outer hair-cells \cite{Nobili1996}, influence of the spiral character of the cochlear \cite{Manoussaki2008}, influence of fluid channel geometry \cite{Parthasarathi2000}. The proposed models are 1-D, 2-D, or 3D using the Finite-Difference Method (FDM) \cite{Neely1981}, Wentzel-Kramers-Brillouin (WKB) approximation\cite{Steele1979}, Euler method \cite{Babbs2011}, Finite-Element Method (FEM) \cite{Kolston1996}, Boundary-Element Methods (BEM) \cite{Parthasarathi2000} or a quasilinear method \cite{Kanis1996}. All these models are stationary. Time-dependent models have been proposed to model otoacoustic emissions \cite{Verhulst2010} or to compare frequency and time domain solutions \cite{Kanis1996}. Using a frequency-domain BM model Ramamoorthy \cite{Ramamoorthy2007}  couples the mechanical BM displacement to the hair bundle (HB) conductance, so that the conductance changes linearly with BM displacement. Only the Ramamoorhy model incorporates the transfer from mechanical to electric5al energy, still not in a time-dependent way. 

So most models are eigenmode analysis and very few consider the transition from mechanical motion to spike excitation. The cochlear model \cite{Bader2015} applied in the present investigation conveys the time-varying synchrony of individual frequency components in the acoustic signal, across the transduction from the mechanical wave on the basilar membrane into electrical spikes within auditory nerve fibers.

Therefore it is expected that the time series of the waveform traveling along the basilar membrane is of importance. In the case of a double-slip waveform each slip will produce a large amplitude peak traveling along the basilar membrane. The excited spike train will contain the spikes at any frequncy induced by the sound. Sill it is expected that the double-slip peak will do the same, causing a secondary spike train. Therefore the time intervals between both peaks, the larger peak of the regular periodicity and that of the double-slip, is expected to be neurally coded.

\section{METHOD}

\subsection{BOWING MACHINE}

The employed bowing machine consists of a pendulum which moves a bow on a strictly straight track. The conversion of a pendulum's inherent circular track into a straight track is by eccentric suspension and by rules of geometry \cite{Mores2015}. While the bow is moved across the string both forces apply strictly orthogonal by means of construction. More precisely, the tractive force, i.e. the force necessary to move the bow and to overcome friction, applies in perpendicular direction to the bow force, i.e. the force that is directed towards the strings, perpendicular to the string-bow plane. Both forces can be set at constant levels to investigate the mentioned various regimes of Helmholtz motion or bifurcation or ´surface´ tones. The forces can also be ramped up or down between two nearby levels to investigate transitions between these regimes and within regimes. Such slight ramping is used here to investigate transitions between different bifurcation regimes and to obtain spike trains containing double-slip peaks with varying timely positions relative to a fundamental period. While bowing, the parameters can be set at certain levels but can also be measured with good precision since the inherent friction of the pendulum is very low and known \cite{Mores2015}. The bow velocity results from applied forces and the stick-slip interaction. In other bowing machines the bow velocity is predefined and controlled by a driving motor \cite{Cronhjort1992} \cite{Galluzzo2014} \cite{Lawgren1980} \cite{Pickering1991} \cite{Schoonderwaldt2008} \cite{Schumacher1996}. Such an approach reveals little about how the stick-slip organizes itself with the given applied forces. Here, in an adaptive fashion, the bow velocity is a direct result not only of applied forces but also of present string motion. Bow velocity is measured with good precision as well.

\subsection{PEAK DETECTION}

The time series, as recorded by the accelerator at the bridge of the cello, is analyzed in the time-domain. Therefore the amplitude peaks of the waveform are detected; and although these peaks repeat each waveform period, some peaks within the fundamental waveform, such as the double-slip peak, might change their position within one period. The development of these peaks from period to period throughout the length of the tone is investigated. 

As discussed above, the cochlear spike train is based on the excitation of neurons mainly by the waveform amplitude peaks. On the other side, the physical process of the double-slip likewise results in a peak in the time series, as does the main tear-off. By tracking the peaks of the time series, especially the double-slip peak, the physical process of the double-slip tear-off is followed.

The tone consists of about 700 periods of the fundamental frequency of 148 Hz. Fig. 2 (upper plot) shows the waveform of one period starting from the main peak P1 where the main tear-off happens, so where the bow is tearing off the string. This results in a vibrating bridge and body motion as recorded by the waveform which decays quite fast as the following two smaller peaks P2 and P3 suggest. Then the double-slip appears as peak DS1 followed by two peaks, again decaying. Here only one peak denoted as DS2 is shown as the other was unstable throughout the 700 periods and therefore could not be tracked consistently throughout the sound and therefore were omitted in the analyses. The lower plot in Fig. 2 shows the same recording for a regular Helmholtz motion. No second tear-off is present in the middle of one period and therefore no DS1 and DS2 are found.

\begin{figure}
	\centering
	\includegraphics[width=0.9\linewidth]{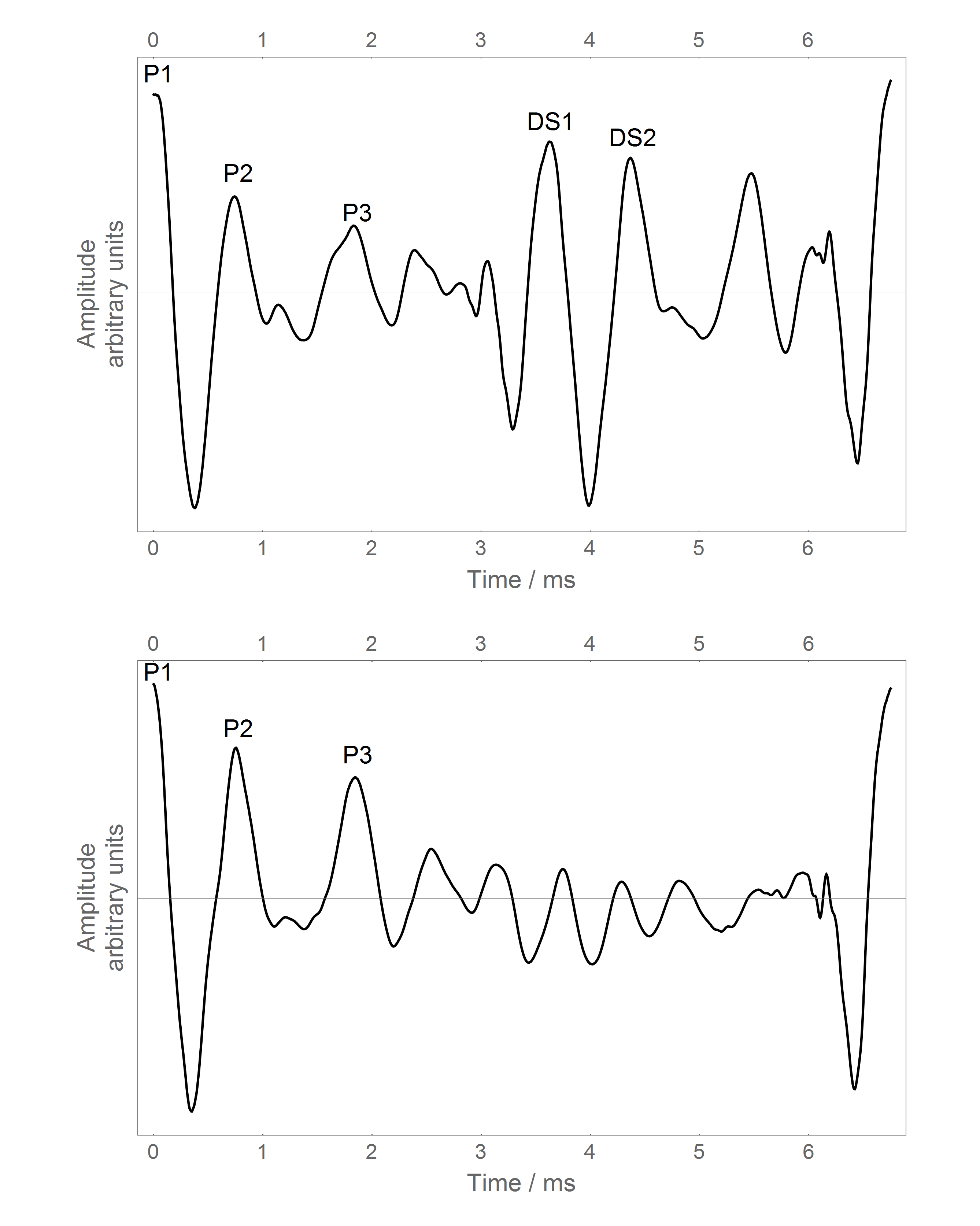}
	\caption{One period of the waveform with tracked peaks for the double-slip sound (top) and a regular Helmholtz motion (bottom). P1: main tear-off bow from string, DS1: double-slip tear-off, P2, P3: peaks following P1, DS2: peak following DS1. The peak between DS1 and DS2 was not always prominent and therefore not tackable throughout the sound and was therefore omitted for analysis. With the normal Helmholtz motion the second peak in the middle is not present and therefore DS1 and DS2 are not there.}
	\label{fig:fig2}
\end{figure}

For all five peaks a tracking algorithm was detecting the time points $z_t^i$  with i = \{P1, P2, P3, DS1, DS2\} of the maximum height of the respective peaks for 700 adjacent periods. From these time points the relative positions of the peaks within each period were calculated as

\begin{equation}
i/ j_t =  (z_t^i - z_t^j)/ (z_{t+1}^{P1} - z_t^{P1}) \ \text{with  i,j = \{P1, P2, P3, DS1, DS2\}} \ .
\end{equation}

So for i=P1 and j = DS1 the relation is labeled P1/DS1 in the following. The time difference $z_t^i - z_t^j$ between two peaks is taken relative to the length of their individual period $(z_{t+1}^{P1} - z_t^{P1}$ they are in, to result in values $0 < i/j < 1$. In case where the relation is $i/j > 1$ the reverse case $j/i < 1$ was used.

The relation of peaks can be associated with harmonic partials. As the main interest is the double-slip peak which appears near the middle of the period, the precise position of this peak is expected to influence the second partial of the sound. If the double-slip is perfectly at P1/DS1 = 0.5 then the second partial might become prominent and the perceived pitch might flip to this higher partial as the waveform is nearly perfectly doubled. Still if P1/DS1 or any other relation such as P3/DS2 is slightly off 0.5 there are two periodicities at the frequency of the second partial, a smaller and a larger one. Therefore the second partial would be bifurcating into two periodicities. We might expect that this leads to the perception of arbitrariness of the sound making it a 'surface' tone. To investigate how this is processed in the ear, a cochlear model is used as discussed in the following section.

\subsection{FINITE-DIFFERENCE TIME DOMAIN (FDTD) COCHLEA MODEL}

The present model has been discussed in detail before \cite{Bader2015}. There, a synchronization of spike phases at the transition between mechanical energy on the basilar membrane (BM) and the excited electrical spike by this energy was shown. The synchronization appears because of the nonlinear transfer condition. This transition is well studied \cite{Hubbard1996} and is implemented using two conditions. A spike at one point X on the BM at one time point $\tau$  is excited if

\begin{enumerate}
\item $u(X-1,\tau) < u(X,\tau) > u(X+1,\tau)$, \\
\item$u(X,\tau-1) < u(X,\tau) > u(X,\tau+1)$ \ .
\end{enumerate}

Condition (1) means a maximum shearing of two hairs at a hair cell as a necessary condition to an opening of the ion channels at this hair cell. This only happens with a positive slope as only then the stereocilia are driven away from each other. With a negative slope the cilia are getting closer and therefore no stress appears at the tip links between them. This corresponds to the rectification process in gammatone filter banks.

 \medskip 
 
Condition (2) is a temporal maximum positive peak of the BM displacement. It is the temporal equivalent to the spatial condition, a maximum acceleration where the tip link between the cell and its neighbouring cells is most active.

 \medskip

A Finite-Difference Time Domain (FDTD) model was implemented as cochlear model. The basic Finite-Difference model was successfully implemented with musical instruments \cite{Bader2013} \cite{Bader2005} and is highly stable and reliable. Finite-Difference methods have previously been used with cochlear models. They have most often been treated as eigenvalue problems \cite{Neely1981}, rather than time-related issues, as proposed above. But when intending time-dependent solutions, FDTD methods are suitable because of stability and realistic frequency representation.

 \medskip

The model assumes the BM to be a rod rather than a membrane as here it is assumed to be 3.5 cm long and only 1 mm wide at the staple and 1.2 at the apex. So a 1-D model is enough when omitting the spiral geometry or the fluid channels in the model \cite{Babbs2011}. The fluid dynamics is neglected because the speed of sound in the lymph, which is about 1500 m/s is much larger than on the BM, which is about 100 m/s at the the staple and decreases fast to about 20 m/s at the apex \cite{Ramamoorthy2010}. Therefore, for a single sinusoidal the same phase is present all along the BM at one time point and the long-wave approximation \cite{deBoer1991} can be used.

 \medskip

Then the 1-D differential equation of the model is linear but inhomogeneous with changing stiffness. In the model, values of \cite{Allen1977} are used with stiffness $K = 2 \times 10^9 e^{-3.4 x} dyn / cm^3$ changing over length x of the BM. The linear density $\mu$  is only slightly changing along the length because of the slight widening of the BM. According to the Allen values the mass is assumed constant over the BM:  $m = 0.05 g / cm^2$.

 \medskip

The damping of the BM is also depending on space. According to the time integration method of the FDTD model a velocity decay at each time integration is applied. Using a sample frequency of $s= 200 kHz$ the Allen damping values $d = 199 - .002857 x \ \frac{dyn\ s}{cm^3}$ are implemented using a velocity damping factor of  $\delta(x) = .995 - 0.00142857 x$.

\section{Results}

\subsection{Waveform peak relations}

As discussed in the section the relations between the peaks of each period are calculated as shown in Fig. 3. The development of the relations are plotted over 700 periods of the tone with fundamental frequency of 148 Hz. The peak relations are labeled within the plot for all ten possible combinations of five tracked peaks. 

\begin{figure}
	\centering
	\includegraphics[width=1.0\linewidth]{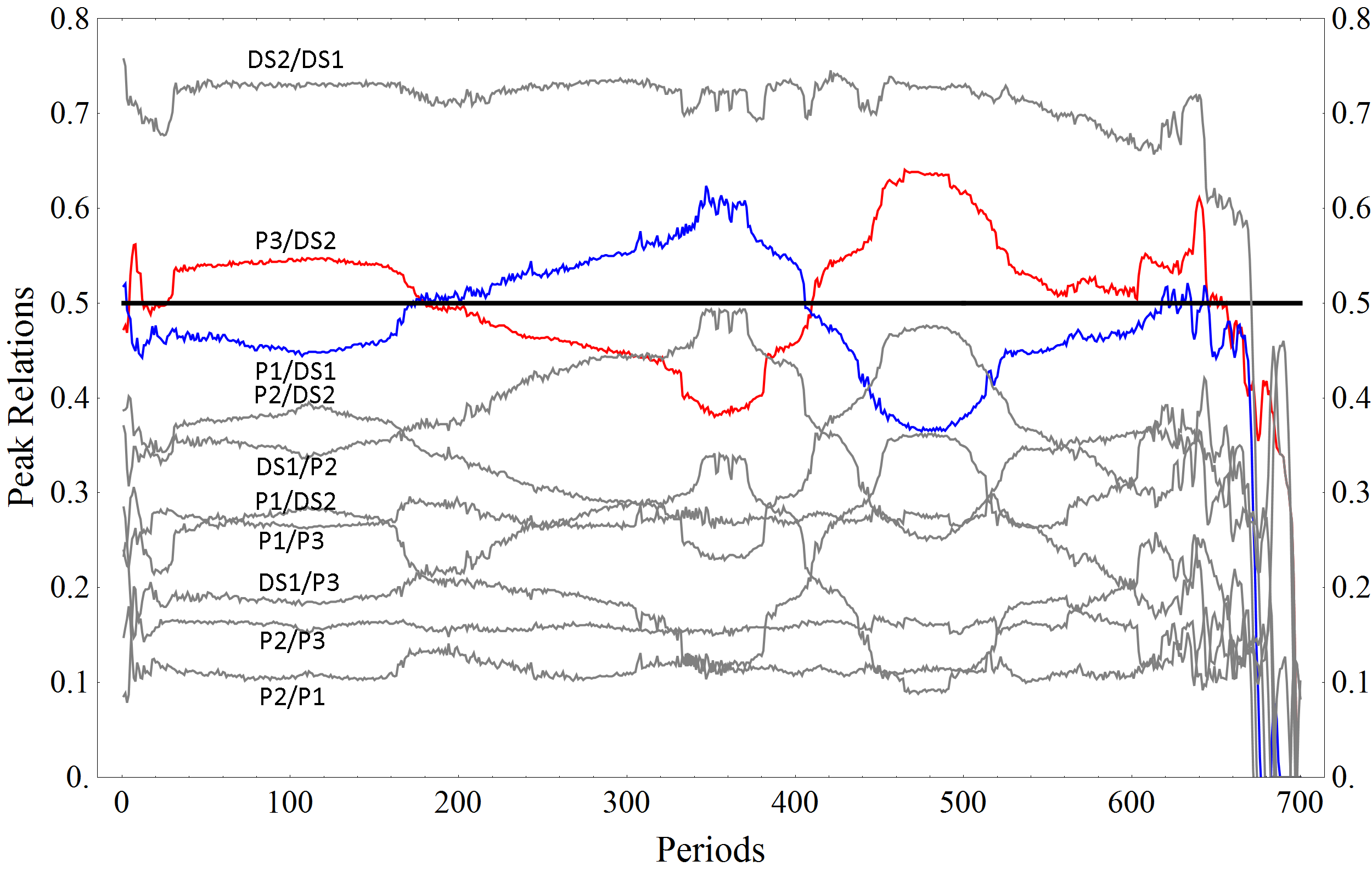}
	\caption{Time series of the peak relations of the waveform of Fig. 2. The two periodicities around 0.5 of the whole periodicity are the relations between the main peak and the double-slip peak (P1/DS1) and the the respective later peaks P3/DS2. The other relations show related developments as P1/DS1 and P3/DS2.}
	\label{fig:fig3}
\end{figure}

The most important relation is the one between the main peak and the double-slip peak P1/DS1, shown in blue. Indeed, the relation is around 0.5 but only at two time points the relation P1/DS1 is precisely 0.5. After the initial transient, the peak is below 0.5, and quite stable, then raises, passing through 0.5 in a quite constant slope. After a peak at about 350 periods it becomes smaller, crossing 0.5 again and after a minimum it increases again towards the end of the tone. This behaviour is followed by a second relation P3/DS2 mirroring nearly perfectly, although not exactly the P1/DS1 case. Both become 0.5 twice and at the same time points. Therefore this relation of 0.5 is indeed bifurcating into two relations in a mirrored fashion.

In terms of perception, listeners reported that the pitch of the played note flipped to the octave right at the two time-points which both periodicities met at a relation of 0.5. There, the two subsections of the main period seem to be so similar that the fundamental frequency seems to have  doubled, then representing a pitch of 296 Hz.

The other relations are mostly below 0.5, with the exception of DS2/DS1, and show similar behaviour with respect to the overall shape as well as the time points where P1/DS1 = 0.5 precisely. As both, the main and the double-slip peak are followed by additional peaks depending on the time points of P1 and DS1, this behaviour is not too surprising. Slight deviations of the other relations from the main P1/DS1 and P3/DS2 relations are caused by the waveform building up slightly different over the duration of the tone. This is most likely caused by the changing bow pressure and velocity discussed below.

\subsection{BIFURCATIONS IN COCHLEOGRAM	}

To estimate the perception of the double-slip, the tone as analyzed in the peak detection algorithm is used as input to the cochlear model discussed above. This results in spikes at the 24 Bark bands at respective time points and with amplitudes according to the traveling wave amplitude causing the respective spike. The time periods between two spikes, the interspike interval (ISI) is summed over all Bark bands and each time interval, resulting in an ISI histogram. Using the ISI as periodicities, each histogram region can be associated with a respective frequency: f = 1/ ISI.

As a result in Fig. 4 the ISI histogram over time for the cello tone is shown. The fundamental frequency at 148 Hz is clearly present throughout the tone with a high consistency. This corresponds to the clear pitch perception of the sound. Still in the frequency region of the second partial around 296 Hz a more complex pattern can be seen. This continues with higher frequencies. As we are mainly interested in the double-slip, which appears at about half of the main period of 148 Hz, the region around 196 Hz is discussed in the following.

\begin{figure}
	\centering
	\includegraphics[width=1.0\linewidth]{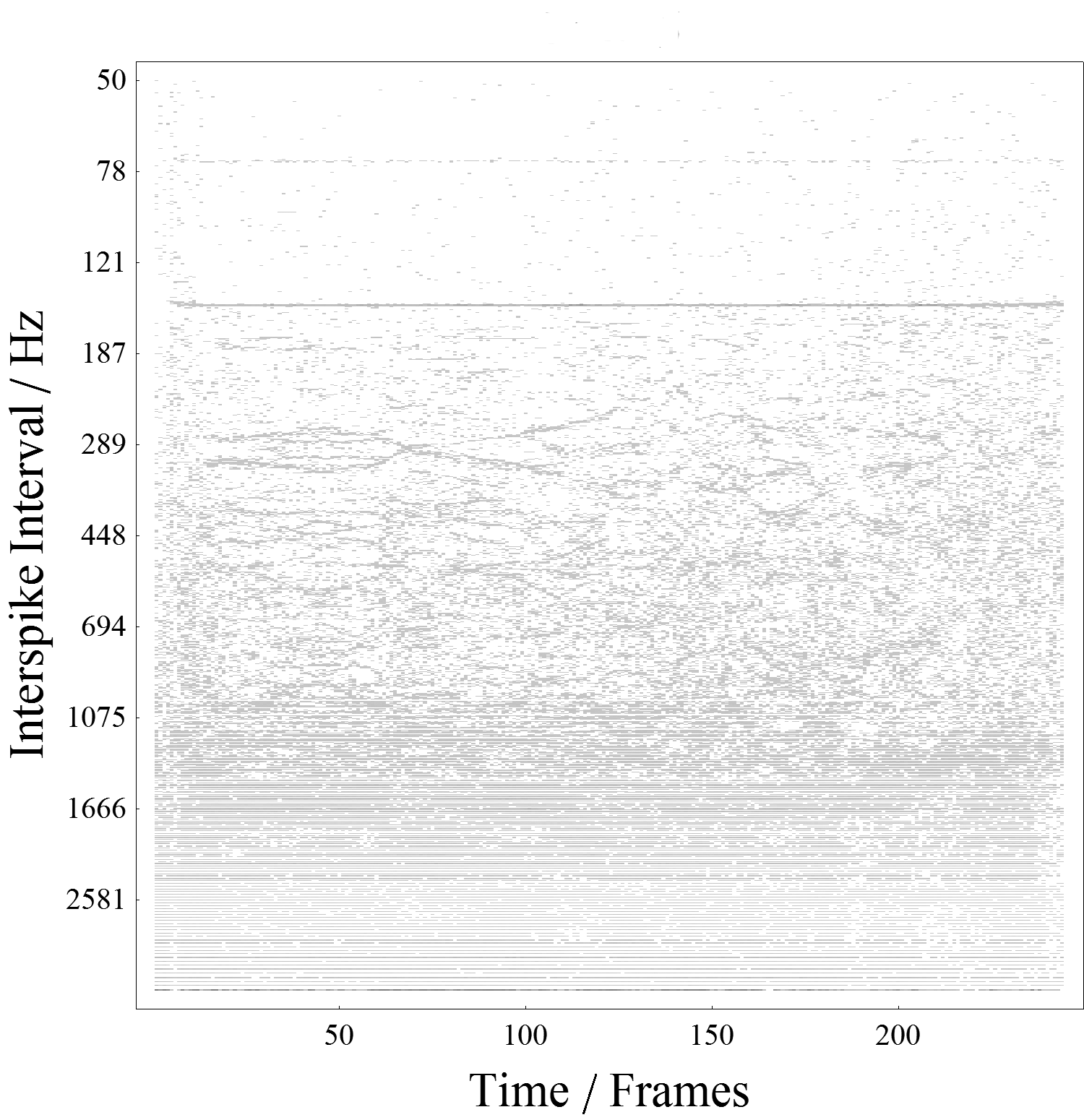}
	\caption{Interspike interval (ISI) histogram as sum over the Bark bands of the cochlear model represented as frequencies over time of the cello tone of 148 Hz fundamental frequency. The fundamental frequency is clearly coded in spikes very precisely and throughout the sound. This corresponds very well with the clear perception of pitch of this tone. Around the frequency of the second partial at 296 Hz, a more complex structure is present which is shown in detail in Fig. 5.}
	\label{fig:fig4}
\end{figure}

Fig. 5 shows this region as an excerpt together with the pressure (blue curve) and the velocity (red curve) of the bowing process. The pressure is increasing until tone onset which is marked at the first black vertical line from the left. Before tone onset the bowing velocity is decreasing. After the tone onset the pressure is decreasing again with a peak at tone onset. Also the velocity is decreasing as expected because of increased friction through the bowing process.

Investigating the ISI historgram, two main trajectories appear within Region I (marked as “I” at the top-left of the graph). As the ISI represents the time intervals between two spikes, there are two periodicities present at the same time around 296 Hz (pressure and velocity labeled in this plot, frequencies labeled in the next plot). These two trajectories slip off in region II into four, showing a typical bifurcation scenario. At the end of region II the four periodicities come together to two again and unite into one within region III only to slip up into two again. At the beginning of region IV two trajectories divide into four again, and in region V a very complex behaviour of periodicities appear.

It is interesting to see that the velocity development is aligned with these regions. At the start of region II it has a minimum, which again holds at the start of region III. Still then it is nearly stable. The pressure on the other side does not show considerable changes at the region boundaries.

\begin{figure}
	\centering
	\includegraphics[width=1.0\linewidth]{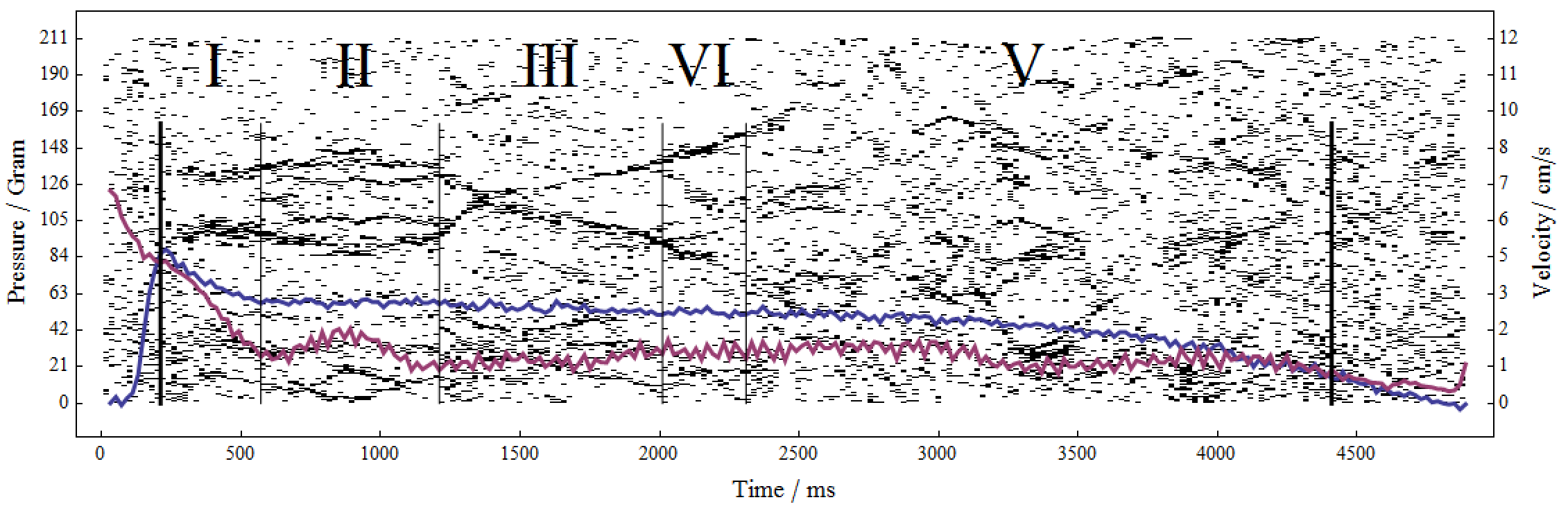}
	\caption{Cochleogram between 210 Hz - 450 Hz of a cello tone with 148 Hz fundamental frequency therefore covering its second partial frequency range bowed with very low bowing pressure (blue curve) and bowing velocity (red curve). The thick vertical lines indicate the tone onset and offset. The thick vertical lines indicate sudden changes between different bifurcation regimes (I): split of partial into two periodicities, (II): split of partial into four periodicities, (III): split of partials again in two periodicities converging into the original frequency of the partial and bifurcating again into two periodicities, (IV): split of partial again into four periodicities, (V): more complex bifurcation regimes.}
	\label{fig:fig5}
\end{figure}

\subsection{CORRESPONDENCE BETWEEN DOUBLE SLIP AND NERVOUS SPIKE RELATIONS}

To compare the cochlear spike train output of the ISI histogram as shown in Fig. 5 with the peak tracking of the waveform, in Fig. 6 both figures are combined. As the double-slip appears at relations around 0.5 where the second partial of the 148 Hz fundamental frequency of the cello tone is expected, only the two relations P1/DS1 and P3/DS2 which indeed are around 0.5 are shown for simplicity.
The cochlear model is able to capture the physical process behind a spike-train, be it a regular Helmholtz motion or a collateral, slightly time varying sequence of intermediate, within-period slips of the bow.
The ISI relations and the peak tracking align very well, both in time as well as in frequency. In terms of time development there were two points in time where a relation of P1/DS1 = P3/DS2 = 0.5 is reached perfectly. Here the bifurcating ISI trajectories join in perfect alignment with the peak tracking at the beginning of region III and at about one third of region V. Also the strength of deviations from a perfect 0.5 relation in the peak tracking corresponds very well with the amount of deviations of the ISI frequency from around 296 Hz. 

Therefore it is clear that the cochlear model is detecting the peak relations of the waveform and represents the sound as ISI relations. This does not disturb the detection of a fundamental frequency at all, which is present as an ISI throughout the tone. 

\begin{figure}
	\centering
	\includegraphics[width=1.0\linewidth]{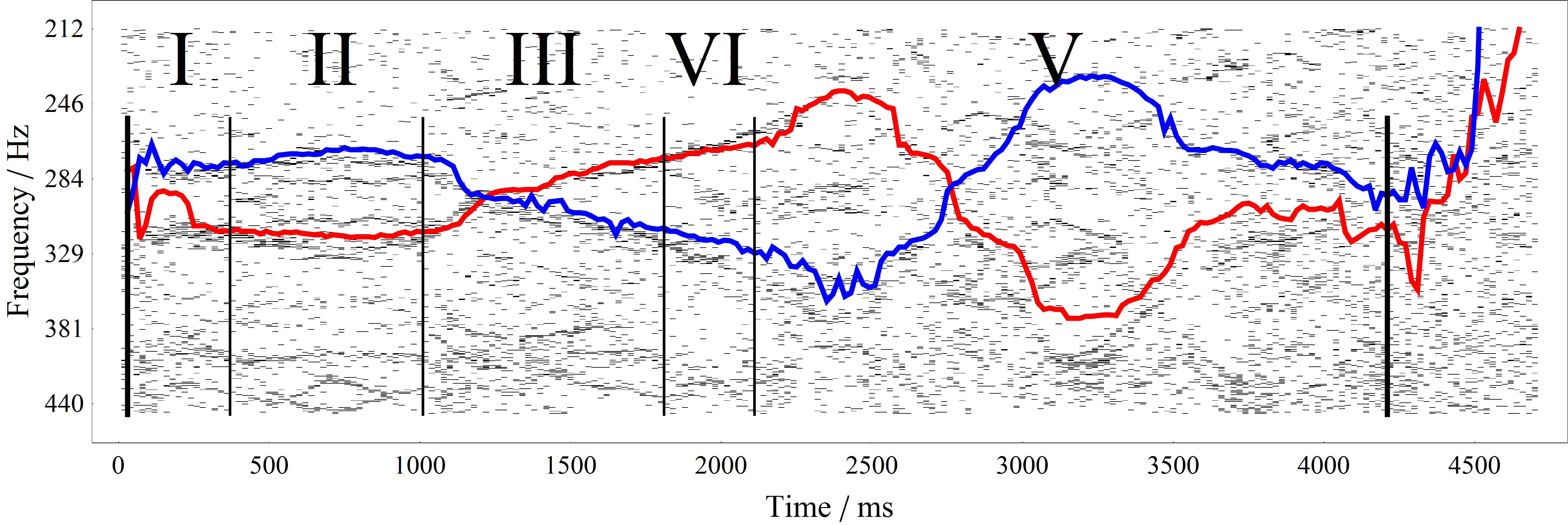}
	\caption{Combination of cochlear model ISI histogram around the frequency of the second partial at 296 Hz and peak tracking of waveform for the cello tone. The trajectories for the P1/DS1 and P3/DS2 plots align very well with the bifurcating periodicities of the cochlear spike output.}
	\label{fig:fig6}
\end{figure}

\begin{figure}
	\centering
	\includegraphics[width=0.9\linewidth]{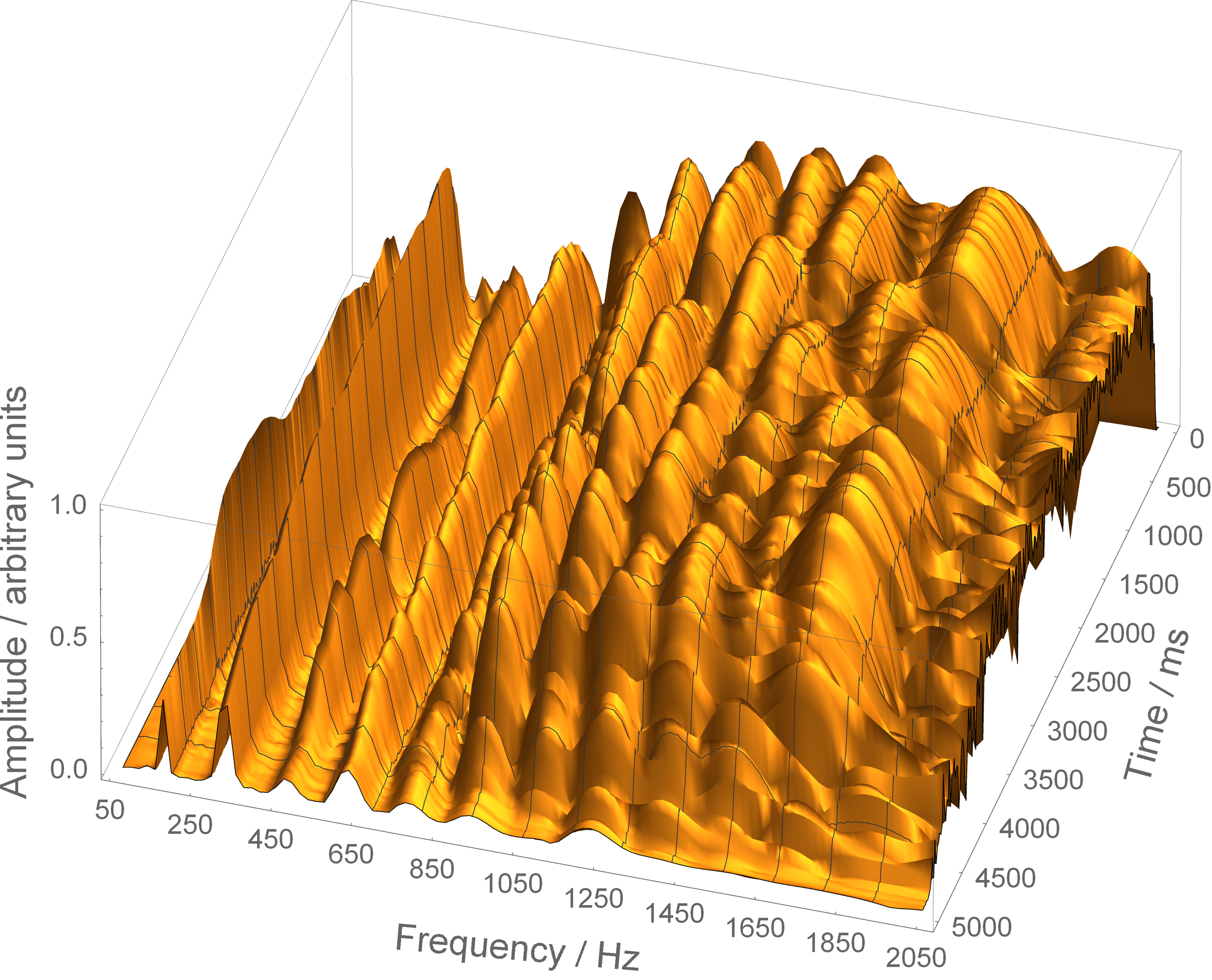}
	\caption{Wavelet representation of the cello tone. The double-slip development is represented as changes in the higher partials. The second partial, where periodicity bifurcation happens, is stable across frequency in this Wavelet transform, and no bifurcations can be seen. The Wavelet transform is not able to represent periodicity bifurcations like the double-slip motion.}
	\label{fig:fig7}
\end{figure}

Comparing the cochlear analysis with a Wavelet-Transform, as shown in Fig. 7 a totally different picture can be seen. The periodicity bifurcation of the double-slip, found in the peak detection as well as in the cochlear model is expected to appear as a bifurcation of the second harmonic. Still the peak of the second harmonic is stable throughout the tone and does not show any doubling of peaks or any development following the changing peak of the double-slip. Of course the Wavelet transform is representing changes in the sound, still these are found in amplitude changes of the higher harmonics above the second partial. This is reasonable from a standpoint of Wavelet transformation, as changes of the time series within one period to another period implies changes in the amplitudes of the spectrum. Still within such a representation, a change of the double-slip peak within a period is not represented as a double peak in the Wavelet spectrum.

\section{Conclusions}

The cochlear model is able to capture the physical process behind a spike-train, be it a regular Helmholtz motion or a collateral, slightly time varying sequence of intermediate, within-period slips of the bow. This is a different, much more reasonable representation than obtained from Wavelet or Fourier transforms. The Wavelet-transform is not able to detect the changing periodicity and the bifurcation of the double-slip at the second partial of the sound. Although it is coding the changing waveform as a change in the higher harmonic content of the sound over time, the physical process underlying the change in the waveform is not detected.  Although with additional feature extraction the double-slip motion might be detected also in the frequency domain using some classification or training algorithm, the cochlear seems to be able to detect the physical process right at the transition from the mechanical wave in the electrical spikes without any further detection operations or signal processing tasks.

Of course the ear does not 'understand' the physics such that there is a bow and a string. Still it represents the physical process as such. This might be caused by the co-evolution of sounding objects and the cochlear and ear. As a conjecture there might have been an adaptation process between the environment physical sound production processes and the adaptation of the ear in such a way that the cochlear detects the physical process most effectively, representing the process in the ISIs without any need to perform further signal processing tasks.


\begin{thebibliography}{99}
	
	\bibitem{Allen1977} J.B. Allen, ``Two-dimensional cochlear fluid model: New results,'' \emph{J. Aocust. Soc. Am.} \textbf{61}, 110-119 (1977).
	
	\bibitem{Babbs2011} Ch.F. Babbs, ``Quantitative Reappraisal of the Helmholtz-Guyton Resonance Theory of Frequency Tuning in the Cochlea,'' \emph{J. of Biophysics} \textbf{ID 435135} (2011), pp. 1--16.
	
	\bibitem{Bader2015} R. Bader, ``Phase synchronization in the cochlear at transition from mechanical waves to electrical spikes,´´ \emph{Chaos} \textbf{25}, 103124 (2015).
	
	\bibitem{Bader2013} R. Bader, \emph{Nonlinearities and Synchronization in Musical Acoustics and Music Psychology,} Springer-Verlag, Berlin, Heidelberg, Current Research in Systematic Musicology, vol. 2,  (2013) pp. 157--284.
	
	\bibitem{Bader2005} R. Bader,
	``Whole geometry Finite-Difference modeling of the
	violin,´´ \emph{Proceedings of the Forum Acusticum 2005}, 629-634
	(2005).
	
	\bibitem{Cariani1999} P. Cariani, ``Temporal Coding of Periodicity Pitch in the Auditory System: An Overview,'' \emph{Neural Plascitity} \textbf{6 (4)}, 142 -172 (1999).
	
	\bibitem{Cariani2001} P. Cariani, ``Temporal Codes, Timing Nets, and Music Perception,'' \emph{J. New Music Research} \textbf{30 (2)}, 107–135  (2001).
	
	\bibitem{Cronhjort1992} A. Cronhjort, ``A computer-controlled bowing machine (MUMS),´´ \emph{STL-QPSR} 33 (1992) pp. 61–-66.
	
	\bibitem{Cremer1985} L. Cremer, \emph{The physics
		of the violin,} MIT Press, Cambridge (1985).
	
	\bibitem{deBoer1991} E. de Boer, ``Auditory physics. Physical principles in hearing theory,'' \emph{Phys. Rep.}, \textbf{203}, 127-229 (1991).
	
	\bibitem{Duffour and Woodhouse 2004a}
	P. Duffour and J. Woodhouse, ``Instability of systems with a
	frictional point contact: Part 1, basic Modelling,`` \emph{J. Sound
		and Vibration} \textbf{271}, 365-390 (2004).
	
	\bibitem{Duffour and Woodhouse 2004b} P. Duffour and Woodhouse, ``Instability of systems with a frictional point contact: Part 2, model extensions´´ \emph{J. of Sound and Vibration} \textbf{271}, 391-410 (2004).
	
	\bibitem{Galluzzo2014} P. M. Galluzzo, and J. Woodhouse, ``High-performance bowing machine tests of bowed-string transients,´´ \emph{Acta Acustica united with Acustica} 100 (2014) pp. 139–-153.
	
	\bibitem{Gueth 1996} W. G\"{u}th, ``A comparison of the
	Raman and the Oscillator Models of String Excitation by Bowing,´´
	\emph{Acustica} \textbf{82}, 169-174 (1996).
	
	\bibitem{Gueth 1995} W. G\"{u}th, \emph{Einf\"{u}hrung
		in die Akustik der Streichinstrumente. [Introduction in the
		acoustics of string instruments.]} Stuttgart 1995.
	
	\bibitem{Gueth 1980} W. G\"{u}th, ``'Ansprache' von
	Streichinstumenten. [Response of string instruments.]´´
	\emph{Asustica} \textbf{46}, 259-267 (1980).
	
	\bibitem{Hubbard1996} A.E. Hubbard and D.C. Mountain, ``Analysis and Synthesis of Cochlear Mechanical Function Using Models,'' in \emph{Auditory Computation}, L. H. Hawkins, T. A. McMullen, A. N. Popper and R. R. Fay, Editors, Springer Handook of Auditory Research, Springer, New York (1996), pp.  62--120.
	
	\bibitem{Hutchings 1997} C. Hutchings (ed.),
	\emph{Research Papers in Violin Acoustics} Vol. I \& II,
	Publication by the Acoustical Society of America, 1997.
	
	\bibitem{Kanis1996} L.J. Kanis and E. de Boer, ``Comparing frequency-domain with time-domain solutions for a locally active nonlinear model of the cochlear,'' \emph{J. Acoust. Soc. Am.} \textbf{100 (4)}, 2543-2546 (1996).
	
	\bibitem{Kimura 1999} M. Kimura, ``How to
	produce subharmonics on the violin,´´ \emph{New Music Research} \textbf{28}, 178--184 (1999).
	
	\bibitem{Kolston1996} P. J. Kolston and J. F. Ashmor, ``Finite element micromechanical modeling of the cochlear in three dimensions,'' \emph{J. Acoust. Soc. Am.} \textbf{99 (1)} 455-467, (1996).
	
	\bibitem{Kumaresana2013} R. Kumaresana, V. K. Peddinti and P. Cariani, ``Synchrony capture filterbank: Auditory-inspired signal processing for tracking individual frequency components in speech,'' \emph{J. Acoust. Soc. Am.} \textbf{133 (6)}, 4290-4310 (2013).
	
	\bibitem{Lawgren1980} B. Lawgren, ``On the motion of bowed violin strings,´´ \emph{Acustica} 44 (1980) pp. 194-–206.
	
	\bibitem{Lyon1996}
	R. Lyon and S. Shamma, ´´Auditory representation of timbre and pitch,'' in \emph{Auditory Computation}, H. L. Hawkins, T. A. McMullen, A. N. Popper and R. R. Fay, Auditory representation of timbre and pitch, Editors, Springer Handbook of Auditory Research, New York (1996), pp. 221--270.
	
	\bibitem{Licklider1956} J.C.R. Licklider, ``Auditory frequency analysis,'' in \emph{Information Theory}, C. Cherry, Editor, (Butterworth, London, 1956), pp. 253--268.
	
	\bibitem{Manoussaki2008} D. Manoussaki, R.S. Chadwick, D.R. Ketten, J. Arruda, E.K. Dimitriadis and J.T. O’Malley, ``The influence of cochlear shape on low-frequency hearing,'' \emph{PNAS} \textbf{105 (16)}, 6162-6166 (2008).
	
	\bibitem{McIntyre et al. 1981} M. E. McIntyre, R. T. Schumacher and J. Woodhouse, ``Aperiodicity in bowed-string motion,´´ \emph{Acustica} \textbf{49}, 13--32 (1981).
	
	\bibitem{McIntyre and Woodhouse 1979} M. E. McIntyre and J. Woodhouse, \emph{Fundamentals of bowed-string dynamics,} \emph{Acustica} \textbf{43}, 93--108 (1979).
	
	\bibitem{Mammano1992} F. Mammano, R. Nobili, ``Biophysics of the cochlear: Linear approximation,'' \emph{J. Acoust. Soc. Am.} \textbf{93 (6)}, 3320-3332 (1992).
	
	\bibitem{Mores2015} R. Mores, ``Precise cello bowing pendulum,´´ \emph{Proceedings of the Third Vienna Talk on Music Acoustics} (2015) 106ff.
	
	\bibitem{Mueller and Lauterborn 1996} G. M\"uller and W. Lauterborn,
	``The bowed string as a nonlinear dynamical system,´´ \emph{Acustica}
	\textbf{8}, 657--664 (1996).
	
	\bibitem{Neely1981} S. T. Neely, ``Finite-Difference solution of a 2-dimensional model of the cochlear,'' \emph{J. Acoust. Soc. Am.} \textbf{69 (5)}, 1363-1393 (1981).
	
	\bibitem{Nobili1996} R. Nobili and F. Mammanou, ``Biophysics of the cochlear II: Stationary Nonlinear phenology,'' \emph{J. Acoust. Soc. Am.} \textbf{99 (4)}, 2244-2255 (1996).
	
	\bibitem{Parthasarathi2000} A.A. Parthasarathi, K. Grosh and A.L. Nuttall, ``Three-dimensional numerical modeling for global cochlear dynamics,'' \emph{J. Acoust. Soc. Am.} \textbf{107 (1)}, 474-485 (2000).
	
	\bibitem{Patterson1995} R.D. Patterson, M.H. Allerhand and Ch. Gigu\`ere, ``Time-domain modeling of peripheral auditory processing: A modular architecture and a software platform,'' \emph{J. Acoust. Soc. Am.} \textbf{98 (4)}, 1890--1894 (1995).
	
	\bibitem{Pickering1991} N. C. Pickering, ``A new light on bow action,´´ \emph{Journal of the Violin Society of America} 11 (1991) pp. 83–-92.
	
	\bibitem{Raman 1918} C. V. Raman, ``On the mechanical theory of
	the vibrations of bowed strings and of musical instruments of the
	violin family, with experimental verification of the results,´´
	\emph{Bulletin 15 of the Indian Association for the Cultivaion of
		Science} \textbf{15}, (1918).
	
	\bibitem{Ramamoorthy2007} S. Ramamoorthy, N. V. Deo and K. Grosh ``A mechano-electro-acoustical model for the cochlear: Response to acoustic stimuli´´ \emph{J. Acoust. Soc. Am.} \textbf{121 (5)}, 2758-2773 (2007).
	
	\bibitem{Ramamoorthy2010} S. Ramamoorthy, D.-J. Zha and A.L. Nuttall, ``The Biophysical Origin of Traveling-Wave Dispersion in the Cochlea,'' \emph{Biophysical Journal} \textbf{99}, 1687–1695 (2010).
	
	\bibitem{Schelleng1973} Schelleng, John C. ``The bowed string and the player,´´ \emph{J. Acoust. Soc. Am.} \textbf{53} 1, 26-41 (1973).
	
	\bibitem{Schoonderwaldt2008} E. Schoonderwaldt, K. Guettler, A. Askenfelt, ``An empirical investigation of bow-force limits in the Schelleng diagram,´´ \emph{Acta Acustica united with Acustica} 94 (2008) pp. 604-–622.
	
	\bibitem{Schouten1962} J.F. Schouten, R. J. Ritsma and B.L. Cardozo, ``Pitch of the residue,'' \emph{J. Acoust. Soc. Am.} \textbf{34}, 1418-1424 (1962).
	
	\bibitem{Schumacher1996} R. T. Schumacher, S. Garoff, ``Bowing with a glass bow,´´ \emph{Journal of the Catgut Acoustical Society} 3 (1996) pp. 9–-17.
	
	\bibitem{Steele1979} C.R. Steele and L.A. Taber, ``Comparison of WKB and finite difference calculations for a two-dimensional cochlear model,'' \emph{J. Acoust. Soc. Am.} \textbf{65}, 1001-1006 (1979).
	
	\bibitem{Verhulst2010} S. Verhulst, T. Dau and Ch.A. Shera, ``Nonlinear time-domain cochlear model for transient stimulation and human otoacoustic emission,'' \emph{J. Acoust. Soc. Am.} \textbf{132 (6)}, 3842-3846 (2012).
	
	\bibitem{Weinrich and Causse 1991} G. Weinrich and R. Causs\'e, ``Elementary stability considerations for bowed-string motion,´´ \emph{J. Acoust. Soc. Am.} \textbf{89}, 887--895 (1991).
	
	\bibitem{Woodhouse and Galluzzo 2004} J. Woodhouse and P. M. Galluzzo, ``The bowed string as we know it today,´´ \emph{Acta Acustica United with Acustica} \textbf{90}, 579--589 (2004).
	
	\bibitem{Woodhouse 2003} J. Woodhouse, ``Bowed String Simulation Using a Thermal Friction Model,´´ \emph{Acta Acustica United with Acustica} \textbf{89}, 355--368 (2003).
	
	\bibitem{Woodhouse and Schumacher 1995} J. Woodhouse and R. T. Schumacher, ``The transient behaviour of models of bowed-string motion,´´ \emph{Chaos} \textbf{5}, 509--523 (1995).
	
	\bibitem{Woodhouse1995} J. Woodhouse, ``Self-sustained musical oscillators ´´ in: M. Hirschberg (ed.): \emph{Mechanics of Musical Insruments} Springer (1995) pp. 185--228.
	
\end{thebibliography}
  \end{document}